\documentclass[10pt]{amsart}
\usepackage{epsfig,epic,amssymb,boxedminipage,multicol}
\usepackage[arrow,curve,all]{xy}
\def\intprod{\mathbin{\hbox to 6pt{%
                 \vrule height0.4pt width5pt depth0pt
                 \kern-.4pt
                 \vrule height6pt width0.4pt depth0pt\hss}}}

\let\hook\intprod
\newtheorem{rmk}{Remark}[section]
\newtheorem{thm}{Theorem}[section]

\newtheorem{cor}[thm]{Corollary}
\newtheorem{prop}[thm]{Proposition}
\newtheorem{defn}[thm]{Definition}
\newcommand{\djb}{J^{1} \pi^\star}             
\newcommand{\isto}{\hspace{5pt} ::\hspace{5pt}} 
\newcommand{\pdjb}{{J^1 \pi^*}^{\Sigma}_{+}} 

\newcounter{alpheqn}
\begin{document}

\title[Ambient Diffeomorphism Symmetries of Embedded Submanifolds... ]
    {Ambient Diffeomorphism Symmetries of Embedded Submanifolds, Multisymplectic BRST and Pseudoholomorphic Embeddings.}
\author{{S.P.Hrabak.}}
\address{Department of Mathematics\\
  King's College London\\ 
  Strand\\
   London WC2R 2LS\\
 England} 
\email{e-mail: shrabak@mth.kcl.ac.uk}
\thanks{Research supported by PPARC}

\subjclass{Primary: 53  Secondary: 70}
\date{August 3rd, 1999}
\begin{abstract}
We describe the multisymplectic analysis of the constraints of the pseudoholomorphic embeddings of Riemann surfaces into  { strictly} almost K\"{a}hler ambient manifolds inherited from  diffeomorphisms of the ambient manifold. The { non-integrability} of the almost K\"{a}hler structure is an { obstruction} to the use of the manifestly non-covariant constraint analysis  of the  Dirac and Bergmann formalism. 

The ambient diffeomorphism deformations  of first-order embedded submanifolds are examined.  We find that the covariant Noether currents, corresponding to the inherited ambient diffeomorphism symmetry, satisfy a non-Abelian deformation algebra, the structure functions being the Cartan structure functions on the ambient manifold.

We define the covariant kinematical phase space of pseudoholomorphic embeddings (the symplectic 2-submanifolds of a symplectic manifold) explicitly as a subbundle of the covariant kinematical phase space of embeddings. The induced algebra of Noether currents satisfies the same algebra as before, the symmetry thus being preserved on this  subclass of embeddings.

The graded multisymplectic manifolds of the  covariant Hamiltonian BRST formalism, developed by  the author, are explicitly constructed for the symmetry of embeddings and pseudoholomorphic embedding. We find that the ``(supersymmetry) multiplet and its dualities'' postulated  by Witten arise naturally as the   local fibre coordinates on the graded covariant phase space of pseudoholomorphic embeddings.															
The BRST algebra for the pseudoholomorphic class is computed. The structure functions implicit in Witten's treatment of the topological sigma model arise as the Cartan structure functions in a Darboux basis on the ambient symplectic manifold. The BRST algebra postulated by Witten is thus derived as  the BRST prolongation of the non-Abelian deformation algebra.

{ This is the first consistent Hamiltonian formulation in the strictly non-integrable case}. The kinematics of pseudoholomorphic embeddings is seen to be an exemplar of a class of constrained dynamical systems requiring,
for their description, a multisymplectic formalism and therefore necessitating the introduction of  a multisymplectic formulation of the classical BRST symmetry.

Appendices are included containing material on the vielbeins employed and on identities of almost K\"ahler structures.

\end{abstract}

\maketitle

\newpage

The study of multisymplectic geometry \cite{Cans,CanG,CanH,Ibort,Martin}  arose in the context of the search for the geometric foundations of classical field theory \cite{Crampin,GIMMSY,Kij1,Kij2,KijSzc1,KijSzc2}. 
In the  paper \cite{jdg1} the author  formulated a homological description of Marsden-Weinstein multisymplectic reduction in the generic context of free and proper group actions on multisymplectic manifolds. The paper which followed  \cite{jdg2}  returned to the specific context of those multisymplectic manifolds which form the geometric foundations of field theory and described the geometric apparatus necessary to formulate the BRST symmetry in the multisymplectic setting.
 
In \cite{jdg2} we applied the covariant Hamiltonian BRST formalism to the field theory of Yang and Mills. We discovered that the covariant Noether currents generating the gauge symmetry were polynomial in the canonical observables. This is rather interesting as these covariant Noether currents are therefore of a simpler form than the corresponding non-covariant Noether currents which depend upon the derivatives of the momenta. However, the simplification of a model with a known Hamiltonian BRST formulation, does not immediately convince one that a covariant Hamiltonian formalism is necessary. What one requires is a model which cannot be formulated in a non-covariant Hamiltonian formalism. We examine such an example in this article. 

In recent studies of fundamental  physics the notion of ``particles'' $\Longleftrightarrow$ ``curves'' as fundamental dynamical objects has been extended  by the notion of ``p-branes'' $\Longleftrightarrow$ ``p-dimensional submanifolds''. Corresponding to the single parameter of curves are several parameters, the intrinsic (local) coordinates on the embedded manifold. There is therefore no obvious choice of single evolutionary parameter, that is, time. By making a choice of a hypersurface one is however able to enforce a distinction and treat the system as an infinite dimensional classical mechanical system. Then so long as any constraints of such a theory are describable in terms of functions of the derivatives of the induced coordinates with respect to the single time parameter of the (n+1) formalism, the generalised Dirac-Bergman Hamiltonian formalism may be applied to their study. That is, the generalised Dirac-Bergman Hamiltonian formalism may be applied if  the constraints are functions, \(\phi\), of the form \(\phi(u^i,p_j)\approx0\) where the \(u^i\) are the generalised coordinates and \(p_j\) are the generalised momenta defined by \(p_i:={\partial \mathcal{L} }{/}{\partial u^j_{,0}}\) where \(\mathcal{L}(x^\alpha,u^i,u^j_{,\alpha})\) is the Lagrangian characterising the dynamical system, \(x^\alpha\) are the independent variables and  \(u^j_{,0}\) is the time derivative of generalised coordinates, viz. the generalised velocity.

However one can imagine a more generic situation  where a defining constraint on a class of embeddings is such that the constraints  depend quintessentialy on the derivatives of the induced coordinates with respect to all of the several independent variables. That is, the constraints are of the form \(\phi(u^i,p^\alpha_j)\approx0\) where the \(u^i\) are the generalised coordinates and \(p^\alpha_j\) are the multimomenta defined by \(p^\alpha_i:={\partial \mathcal{L} }{/}{\partial u^j_{,\alpha}}\) where \(\mathcal{L}(x^\alpha,u^i,u^j_{,\alpha})\) is the Lagrangian characterising the dynamical system, \(x^\alpha\) are the independent variables and  \(u^j_{,\alpha}\) are the  derivatives of generalised coordinates with respect to all of the independent variables, viz. the multivelocity. 

An example of such a constraint is that defining the embedding, $\phi$, of a pseudoholomorphic curve \((\Sigma,\epsilon)\) into a strictly almost K\"{a}hler manifold \((\text{M},\omega,\text{J})\), \(\epsilon\) being the complex structure on \(\Sigma\), \(\omega\) being the symplectic structure on M, and  J being the strictly almost K\"{a}hler structure on M. 
\begin{figure}[h]
\[
\xymatrix{
TM \ar[r]^J \ar[d]^{\phi_*} & TM \ar[d]^{\phi_*}\\
T\Sigma \ar[r]^\epsilon& T\Sigma
}
\]
\caption{Pseudoholomorphic embeddings.}\label{Fg}
\end{figure}
The  commutative diagram in Figure \ref{Fg} defines  pseudoholomorphicity, where \(\phi_*\) is the push-forward induced by the embedding \(\phi\).

The strictly almost complex structure (this means an almost complex but not complex structure) arises by virtue of the fact that M is symplectic. For every non degenerate two form on a symplectic manifold $M$ there exists a compatible almost complex structure $J$ \cite{dusa}. Every symplectic manifold therefore carries a compatible almost complex structure. As a consequence of the closure of the symplectic 2-form the compatible almost complex structure on a symplectic manifold with a given Riemannian metric is in fact  an  almost K\"{a}hler structure \cite{Goldberg}. Examples of symplectic manifolds with an  almost K\"{a}hler structure which is not K\"{a}hler have been constructed in \cite{McDuff,Thursden,Gompf}. What is to follow is not therefore a vacuous  exercise.

Let the local real analytic coordinates on \(\Sigma\) be \((x^\alpha)_{\alpha=1,2}\) and the real analytic coordinates on M be \((u^i)_{i=1,\cdots,\text{dim M}}\). The {\it pseudoholomorphicity embedding condition}, expressed in terms of the {\it multimomenta}, takes the form\footnote{ This may be explicitly verified by taking the standard  form of the  constraint for pseudoholomorphic mappings, viz. \( u^i_{,\beta}\epsilon^\beta_\alpha=J^i_j u^j_{,\alpha}\)  and Legendre transforming the usual constraint  written in terms of the multivelocities. We shall not introduce the action in this paper as we emphasise here the power of the natural geometric structures of the multisymplectic formalism in determining  dynamical structures.}:
\begin{equation}
\epsilon^\alpha_\beta p^\beta_i=J_i^j p^\alpha_j.\label{phe}
\end{equation}
Note that we place emphasis here on the case of a {\it strictly} almost complex structure J on the ambient  manifold. It is only in this case that the author contends that the multisymplectic formalism is {\it indispensable}. For let us suppose that  the almost complex structure is integrable, i.e. we have an integrable K\"{a}hler structure. Then there exist {\it global} (anti-) holomorphic coordinates on the ambient manifold: \((u^a,u^{a*})\). The pseudoholomorphicity condition then becomes \( u^a_{,\beta}\epsilon^\beta_\alpha=J^a_b u^b_{,\alpha}=i\delta^a_b u^b_{,\alpha}=iu^a_{,\alpha}\). This equation tells us that the embedding is holomorphic so that the constraint is expressible in terms of the vanishing of a single generalised velocity, viz:\[u^a_{,\Bar{z}}\approx0\]
where \((z,\Bar{z})\) are the global (anti-) holomorphic coordinates on the embedded Riemann surface. As a consequence one may use the Dirac-Bergman formalism in the integrable case, as has been done in the literature\footnote{Note that the primary interest in the theoretical physics community has been with the integrable case. This is because the ambient manifold was invariably chosen, due to  physical grounds related to string theory, to be a Calabi-Yau manifold, which possesses an integrable K\"{a}hler structure. The non-integrability assumes importance here  as an exemplar of a field theory requiring  multisymplectic geometry for its Hamiltonian description.}. In the {\it non-integrable case} one cannot write the pseudoholomorphic constraint equation in terms of a single momentum. This prevents one using the usual manifestly non-covariant Hamiltonian formalism. The multisymplectic formalism is well suited to this type of constraint. The non-integrability of the almost K\"{a}hler structure is seen to be  an obstruction to the use of the Dirac-Bergman formalism.

In this article we study the particularly illuminating example of pseudoholomorphic mappings.  We consider the kinematics of pseudoholomorphic mappings into a strictly almost K\"{a}hler manifold. We are able to elucidate the origin of the non-Abelian deformation algebra corresponding to the infinitesimal deformation of the pseudoholomorphic maps into  almost K\"{a}hler manifolds first implicitly introduced by Witten in \cite{TSM}. In the only literature on the Hamiltonian BFV formulation following Witten's paper  one finds the assumption that the deformation algebra is Abelian, which is one reason why the BRST algebra was postulated, see \cite{Igarashi}, rather than derived from an underlying symmetry. The deformation algebra is in fact only Abelian locally. The global symmetry is inherited from  the diffeomorphisms of the ambient manifold and as such is non-Abelian.   

In this paper the ambient diffeomorphism deformations  of embedded submanifolds are examined. We find that the covariant Noether currents satisfy a non-Abelian deformation algebra, the structure functions being the Cartan structure functions on the ambient manifold. We define the covariant kinematical phase space of pseudoholomorphic embeddings explicitly as a subbundle of the covariant kinematical phase space of embeddings. The induced algebra of Noether currents satisfies the same algebra as before, the symmetry thus being preserved on this  subclass of embeddings. The graded multisymplectic manifolds of the authors covariant Hamiltonian BRST formalism developed in \cite{jdg2} are explicitly constructed for the symmetry of embeddings and pseudoholomorphic embedding. We find that the ``(supersymmetry) multiplet and its dualities'' introduced  by Witten arise naturally as the   local fibre coordinates on the graded covariant phase space of pseudoholomorphic embeddings.	The BRST algebra for the pseudoholomorphic class is computed. The structure functions implicit in Witten's treatment of the topological sigma model arise as the Cartan structure functions in Darboux coordinates on the ambient symplectic manifold. The BRST algebra postulated by Witten in \cite{TSM} is thus derived as  the BRST prolongation of the non-Abelian deformation algebra.
This is the first consistent Hamiltonian formulation in the strictly almost K\"{a}hler case in the literature.

Nowhere in this paper will  we  introduce an action. The kinematics is described entirely in terms of natural geometric structures. This suggests that the multisymplectic  formalism might be of utility in studying those models for which an action is either not known or argued not to exist. Such models have  arisen in the study of brane dynamics, in particular there are arguments that suggest that an action for the the M-theory five brane should not exist \cite{Witten,Peter}. Much recent progress in studying such models has been made using a purely geometrical formalism known as the superembedding approach \cite{Howe,Oliver}. Furthermore calibrated submanifolds play a central role in these models \cite{Gauntlett,OF1,OF2,OF3}. Analogous to the pseudoholomorphicity embedding condition, these calibrated submanifolds possess a set of equations known as the Monge-Amp\'ere equations. This area might be a particularly fertile ground in which to seek more examples of dynamical constrained systems which, like the case of the pseudoholomorphic mappings into strictly almost K\"{a}hler manifolds, require a covariant Hamiltonian formalism.
For example, Mark Gross \cite{Gross} has suggested that the study of  pseudo-holomorphic curves on  symplectic manifolds with an almost complex structure is analogous to the study of special Lagrangian submanifolds (i.e. with the associated Monge-Amp\'ere equations) on a complex manifold with an almost symplectic structure. 

Another motivation for studying the multisymplectic formulation of field theory is the hope that a quantisation framework might arise which would be inherently free of the need for renormalisation. The same underlying philosophy was  already evident in Dirac's attempts to reformulate classical electrodynamics, both  as a two dimensional field theory and in terms of the dynamics of streams of matter \cite{Dirac1,Dirac2,Dirac3,Dirac4,Dirac5}. Recent progress has been made in developing a quantised version of the covariant Hamiltonian formalism, see \cite{kan4,kan5}. 

An outline of this paper is as follows:\begin{itemize}
\item \S1: We recall here Witten's postulates and  heuristic construction  of the topological sigma model.

\item \S2: The multisymplectic formulation of the kinematics of first order embedded submanifolds is described. The ambient diffeomorphism deformation symmetries  of first order embedded submanifolds are examined. We find that the covariant Noether currents satisfy a non-Abelian deformation algebra, the structure functions being the Cartan structure functions on the ambient manifold.

\item \S3: We define the covariant kinematical phase space of pseudoholomorphic embeddings explicitly as a subbundle of the covariant kinematical phase space of embeddings.

\item \S4: The  algebra of Noether currents induced upon the subbundle of pseudoholomorphic embeddings is shown to satisfy the same algebra as the original symmetry, the symmetry thus being preserved on this  subclass of embeddings.

\item \S5: The graded multisymplectic manifolds of the authors covariant Hamiltonian BRST formalism are explicitly constructed for the symmetry of embeddings and pseudoholomorphic embedding. 

\item \S6 In \cite{TSM} Witten postulated a certain nilpotent symmetry of a (supersymmetry) multiplet. We find that the ``(supersymmetry) multiplet and its dualities'' introduced postulated by Witten arise naturally from the   local fibre coordinates on the graded covariant phase space of pseudoholomorphic embeddings. The nilpotent symmetry which Witten postulated is shown to be the covariant BRST symmetry arising as the BRST prolongation (see \cite{jdg2}) of the non-Abelian deformation algebra derived in \S4. The BRST algebra for the pseudoholomorphic class is computed. 

\item \S7 The structure functions implicit in Witten's treatment of the topological sigma model are shown to arise as the Cartan structure functions in Darboux basis on the ambient symplectic manifold. The BRST algebra postulated by Witten is thus derived as  the BRST prolongation of the non-Abelian deformation algebra.

\item Appendices are included containing material on the vielbeins employed and on identities of almost K\"ahler structures. 
\end{itemize}
Throughout this paper we use the vertical formalism of Kanatchikov, a brief review may be found in \cite{jdg2}. For further details  see Kanatchikov's original papers \cite{kan1,kan2,kan3}. We freely use the material contained in the papers \cite{jdg1,jdg2} throughout this paper, and therefore assume that the reader has these references to hand. This is a unavoidable consequence of the fact that this paper is a direct application of the structures expounded in \cite{jdg1,jdg2}. To have made this paper self contained would have involved an unacceptable amount of repetition.

\section*{Acknowledgements} The author would like to express his thanks to Jim Stasheff and Alice Rogers for useful communications on a earlier version of this preprint.

\section{Witten's postulates.}

In this section we shall describe to the reader the formulation of the model and the postulated algebra  as originally given by Witten \cite{TSM}. Although the arguments are of the type common in the supersymmetry literature they are heuristic in nature, and from the point of view taken in this paper, mask the true origin of the ground from which the structures Witten defined actually  arise. The aim of this paper is to elucidate this ground.

As explained in the foreword the basic idea is to consider embeddings of a Riemann surface $\Sigma$ into a symplectic manifold $M$, viz.
\begin{equation}
\phi:\Sigma \longrightarrow M \isto \phi:x^\alpha\longrightarrow u^i(x^\alpha).
\end{equation}
Witten then defined a multiplet by introducing  the following field structures:
\begin{itemize}
\item $u^i(x^\alpha)$: the local embedding coordinates.
\item $\chi^i(x^\alpha)$: an anticommuting section of $\phi^*(TM)$.
\item $\varrho_\beta^i$: an anticommuting field ($\beta$=1,2  a tangent index to $\Sigma$, and $i=1,\dots,m$ runs over a basis of $\phi^*(TM)$), obeying the ``self-duality'' constraint: 
\begin{equation}
\varrho_\beta^i\epsilon^\beta_\alpha =\varrho_\alpha^j \text{J}_j^i \label{duality1}\tag{D1}
\end{equation}
\item $H^{\alpha i}$: a commuting field ($\beta$=1,2  a tangent index to $\Sigma$, and $i=1,\dots,m$ runs over a basis of $\phi^*(TM)$), obeying the ``self-duality'' constraint: 
\begin{equation}
H^{\beta i}\epsilon_{\beta}^{ \alpha} =H^{\alpha j}\text{J}_j^i \label{duality2}\tag{D2}
\end{equation}
\end{itemize}
Clearly the ``self-duality'' constraints where inspired by the pseudoholomorphic embedding condition (\ref{phe}).

Then Witten ``postulated'' the following ``fermionic'' transformation laws:
\begin{equation}
  \delta u^i(x^\alpha) = \varepsilon \chi^i(x^\alpha) \label{s2} \tag{P1}
\end{equation}
\begin{equation}
 \delta \chi^i(x^\alpha)=0 \label{s1}\tag{P2}
\end{equation}
The two transformations (\ref{s1}) and (\ref{s2}) are the usual nilpotent supersymmetry. 
\begin{equation}
\delta \varrho_\alpha^i =\varepsilon (H_{\alpha}^i   + \tfrac{1}{2}\epsilon_{\alpha \beta}(D_kJ^i_j)\chi^k\varrho^{\beta j}) - \varepsilon \Gamma^i_{jk}\chi^j\varrho^{k}_\alpha\label{s3}\tag{P3}
\end{equation}
The rather complicated structure of the transformation (\ref{s3}) is explained by Witten as follows: The term proportional to $D_kJ^i_j$ is needed for ``consistency'' with the duality relation (\ref{duality1}). The non-covariant term is needed for covariance under reparametrisation under the $u^i$. The transformation (\ref{s3}) is then regarded as defining $ H_{\alpha}^i$ so that any terms which might be added to the right hand side of (\ref{s3}) could be absorbed in the redefinition of $H$.
Finally Witten postulates the ``unpromising-looking formula'':
\begin{equation}
\begin{split}
\delta H^{\alpha i}&=-\tfrac{\varepsilon}{4}\chi^k\chi^l(\text{R}_{kl}{}^i{}_t +\text{R}_{klrs}J^{ri}J^s_t )\varrho^{\alpha t} \\
&+\varepsilon\epsilon_{\alpha \beta}(D_kJ^i_j)\chi^kH^{\beta j} \\
&-\tfrac{\varepsilon}{4}(\chi^kD_kJ^i_s)(\chi^lD_lJ^s_t)\varrho^{\alpha  t} \\
&-\varepsilon \Gamma^i_{jk}\chi^j H^{ \alpha k}\label{s4}
\end{split}\tag{P4}
\end{equation}
After a rather lengthy calculation, one finds that $\delta^2  H^{\alpha i}=0$.
Witten then goes on to construct an action as a $\delta$-variation of some functional of the fields. After the construction of this action Witten then derives the Noether current corresponding to the $\delta$-variation symmetry, viz.
\begin{equation}
{\mathcal J}^\alpha =g_{ij}H^{\alpha i}\chi^j + J^{is}\varrho_s^\alpha D_kJ_{ij}\chi^k\chi^j \label{cnc}
\end{equation}
So summarising, in  \cite{TSM} Witten:
\begin{itemize}
\item Introduces the field multiplet and certain constraints inspired by the psuodohomorphic embedding condition.
\item Postulates a Grassmann-odd Symmetry of this multiplet.
\end{itemize}
In this paper we shall elucidate the ground from which the definitions and postulates of Witten arise. We identify the Noether current (\ref{cnc}), an  ``a posteriori derived object'' with the ``canonical'' Grassmann-odd momentum observable whose adjoint action (in the sense of the Leibniz bracket), generates the action induced by the prolongation to the graded covariant phase space of the original morphism of the configuration bundle. We shall see that all of the above structures are natural geometric structures on the graded covariant phase space of the authors covariant Hamiltonian BRST formalism as expounded in \cite{jdg2}. In \cite{TSM} Witten also formally introduces a generator Q which is said to generate the above transformations (\ref{s2})-(\ref{s4}) by the adjoint action of bracket $\{\text{Q},\cdot\}$. No explicit details of this formally introduced object are ever given. The notion that these transformations could be the transformations generated by a Hamiltonian BRST charge was taken up in \cite{Igarashi}. But here the authors do not derive the above transformations but simply restate the postulates in the BFV language. In this paper we give substance to these formal statements.

\section{Ambient diffeomorphism deformations of first-order embedded submanifolds}

In the multisymplectic formulation of field theories the space of configurations of  classical field theories is defined to be a bundle over some parameter space the sections of which are the fields of interest. In the case of embeddings the ``fields'' are the sections of the bundle: 
\[
\xymatrix{ M \times N \ar[d]^\pi \\ N \ar@/_15pt/[u]_{\phi}
}
\]
and as such are the graphs of the embeddings of N into M. Note that not all the fields of physics are sections of vector bundles, for here we have a bundle who's fibres are differentiable manifolds. The local coordinates on M are  $(u^i)_{i=1,\cdots,m}$ and those on N be $(y^\mu)_{\mu=1,\cdots,n}$.

The covariant kinematical canonical phase space corresponding to ``first order embeddings''\footnote{``first order embeddings''$:=$ ``embeddings specified up to first order contact and no higher''.}, is the vertical covariant canonical phase space bundle (e.g. as described in \cite{jdg2}). The total space of this bundle is the affine dual of the first jet bundle, $\djb$,  (see \cite{Crampin,jdg2}), the base space is the to be embedded submanifold N. The local coordinates on the  covariant canonical phase space bundle, adapted to the bundle $\pi$ are $(y^\mu,u^i,p_\mu^i)$. Note that because the configuration bundle is trivial we may choose the trivial connection $\mathfrak{A}$=0 in this case the  covariant canonical phase space bundle $\djb$ is endowed with the following multisymplectic (n+1)-form \cite{kan1}:
\begin{equation}
\Omega^V= du^i \wedge dp_i^\mu \wedge d^{n-1} y_\mu
\end{equation}
where $d^n y$ is the volume form on N and $d^{n-1} y_\mu:=\tfrac{\partial}{\partial y^\mu}\hook d^n y$.  We then have:
{\defn
The {\it vertical covariant canonical phase space of first order embeddings} is the quadruple $(\djb, \tau^o , N,{\Omega^V})$.
}

In the multisymplectic formalism symmetries of field theories arise as   morphisms of the configuration bundle which prolong to mutisymplectomorphisms of the covariant  canonical phase space. We are interested here in the possible symmetries of an embedded submanifold inherited from the ambient space. It will be this symmetry which will underly the postulated transformations (\ref{s2})-(\ref{s4}) of Witten.

We wish to consider the covariant multimomentum mapping arising from the vertical bundle morphism of the configuration bundle:
\[
\xymatrix{M\times N  \ar[rr]^{\text{Diff(M)}\times\text{Identity}} \ar[dr]^\pi& & \ar[ld]_\pi  M\times N\\
& N
}
\] 
given by the action of Diff(M) on the fibres $\pi^{-1}(u)\cong M$. 
The Lie algebra corresponding to the infinite dimensional group of diffeomorphisms of M is denoted D(M) and is isomorphic to the space of all smooth vector fields on M, the corresponding diffeomorphisms being generated by the 1-parameter flows generated by the elements of D(M). 

We wish to construct covariant Noether currents which will encode the invariance of our embeddings under this ambient diffeomorphism symmetry. The key idea will be to make use of the fact that we may re-express the notion of diffeomorphism invariance in terms of vertical automorphisms of the frame bundle. 

Consider the automorphisms, Aut$\mathcal{F}$(M), of the frame bundle ($\mathcal{F}$(M)$,\pi^\mathcal{F}$,M). $\phi\in$Aut$\mathcal{F}$(M) iff $\phi$ is a diffeomorphism: 
\begin{equation} 
\phi:\mathcal{F}\text{(M)}\longrightarrow\mathcal{F}\text{(M)} \text{ s.t. }
\phi(\mathfrak{f}\cdot g)=\phi(\mathfrak{f})\cdot g 
\end{equation}
for each $\mathfrak{f}\in$$\mathcal{F}$(M) and for all $g\in$G, where G is the structure group of $\mathcal{F}$(M). Such a $\phi$ satisfies $\pi^\mathcal{F}\circ\phi(\mathfrak{f}\cdot g)=\pi^\mathcal{F}\circ\phi(\mathfrak{f})$ and hence determines a diffeomorphism  $\bar{\phi}$ of M defined by
\begin{equation} 
\bar{\phi}(\pi^\mathcal{F}{\mathfrak f}):=\pi^\mathcal{F}\circ\phi({\mathfrak f}).
\end{equation}
One therefore has a homomorphism
\(j:\text{Aut} {\mathcal{F}}(M)\longrightarrow \text{Diff(M)}\). The kernel of this homomorphism, Gau$\mathcal{F}$(M), is called the {\it gauge group of the second kind}, the structure group G being known as the {\it gauge  group of the first kind}. One then clearly has the following short exact sequence of groups:
\begin{equation} 
\xymatrix{0 \ar[r] & \text{Gau}{\mathcal F}\text{(M)}\ar[r]^i &\text{Aut} {\mathcal{F}}\text{(M)}\ar[r]^j & \text{Diff(M)}\ar[r] & 0
 } \label{sesg}
\end{equation}
Correspondingly one has (see \cite{Guill}) the following short exact sequence of Lie algebras: 
\begin{equation} 
\xymatrix{0 \ar[r] & \text{gau}{\mathcal F}\text{(M)}\ar[r]^i &\text{aut} {\mathcal{F}}\text{(M)}\ar[r]^j & \text{D(M)}\ar[r] & 0
 } \label{sesl}
\end{equation}
where:
\begin{itemize} 
\item $\text{gau}{\mathcal F}$ is the Lie algebra of G-invariant vertical vector fields,
\item $\text{aut} {\mathcal{F}}\text{(M)} $ is the Lie algebra of G-invariant  vector fields,
\item $\text{D(M)} $ the algebra of smooth vector fields on M.
\end{itemize} 
 Let $\mathfrak{s}$ be any section of $\mathcal{F}$(M) representing a Cartan moving frame, that is:
\begin{equation} 
\mathfrak{s}:M \longrightarrow \text{L}\mathcal{F}\text{(M)} \text{ :: }
\mathfrak{s}_A:u^i \longrightarrow {\Hat{\theta}_A}:=E_A^i \frac{\partial}{\partial u^i}
\end{equation}
where the $E_A^i$ are a (not orthonormal) vielbien basis (see Appendix I), and $A=1,\cdots,m$ is an anholonomic index.
Observe that a choice of moving frame (the section $\mathfrak{s}_A$) thus corresponds to a choice of equivalence class in the quotient D(M)$\cong\tfrac{\text{aut} {\mathcal{F}}\text{(M)}}{\text{gau}{\mathcal F}\text{(M)}}$.
\begin{figure}[h]
\[
\input{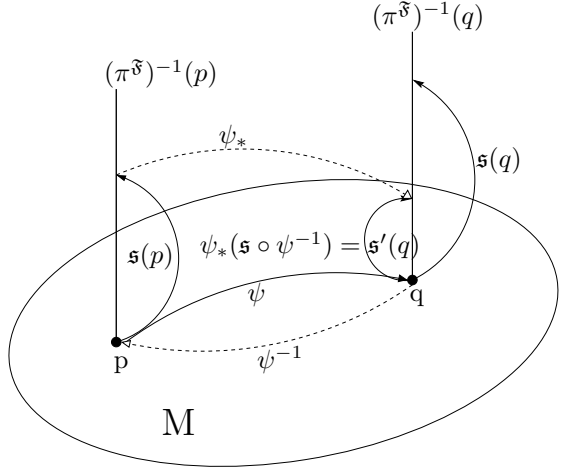}\]\caption{Diff(M) as global  gauge transformations of the second kind.}\label{CT}
\end{figure}

The salient point is that the invariance under Diff(M) can be re-expressed in the form of a gauge invariance under a subgroup of vertical automorphisms of the linear frame bundle achieved by mapping all infinitesimal diffeomorphisms in M into infinitesimal gauge transformations of the second kind that take place along the fibres of  $\mathcal{F}\text{(M)}$, (see \cite{Prugo} for a more detailed discussion). 

For any $\psi\in$Diff(M) we can define a corresponding Cartan transformation, $ \text{D}(\psi)$,  by:
\begin{equation} 
\text{D}(\psi):\mathfrak{s}=({\Hat{\theta}_A})\mapsto\mathfrak{s'}:=
\psi_*(\mathfrak{s}\circ\psi^{-1})
\end{equation}
to another frame $\mathfrak{s'}$ with the same domain as that of $\mathfrak{s}$. Covering $\mathcal{F}\text{(M)}$ with a bundle atlas of local trivialisations associated with local sections of $\mathcal{F}\text{(M)}$, we can thus associate to each $\psi\in$Diff(M) a vertical automorphism $ \text{D}(\psi)$ \cite{Prugo}. This gives a isomorphism between Diff(M) and a subgroup G$_A$(Diff(M)) of the vertical automorphisms of $\mathcal{F}\text{(M)}$. Thus for each $ \text{D}(\psi)\in$G$_A$(Diff(M)) the action of the diffeomorphism $\psi\in$Diff(M) has been transformed into a gauge transformation of the second kind mapping the section $\mathfrak{s}$ into another section $\mathfrak{s'}$, see Figure \ref{CT}.

Given a choice of  moving frame, ${\Hat{\theta}_A}$, any other can be obtained from it by virtue of a gauge transformation of the second kind. In this way one can generate from any arbitrary equivalence class in  
\begin{equation}  \text{D(M)}\cong\frac{\text{aut} {\mathcal{F}}\text{(M)}}{\text{gau}{\mathcal F}\text{(M)}}\end{equation} all others by the action of G$_A$(Diff(M)).

It is thus sufficient to show that our embedded submanifolds are invariant under the 1-parameter group of diffeomorphisms generated by an arbitrary choice of moving frame, or equivalence class in D(M)$\cong\tfrac{\text{aut} {\mathcal{F}}\text{(M)}}{\text{gau}{\mathcal F}\text{(M)}}$.
Let M be a Riemannian manifold with metric g, of signature (p,q). Then we may reduce the structure group of the frame bundle to {\bf SO}(p,q), and thus obtain the orthonormal frame bundle L$\mathcal{F}$(M). We shall henceforth work with an orthonormal frame.
We are free to choose the Levi-Civita connection on the  the Riemannian manifold M.
A Cartan  moving frame satisfies the following algebra:
\begin{equation}
[\Hat{\theta}_A,\Hat{\theta}_B]=C(u)^C_{AB} \Hat{\theta}_C
\end{equation}
where [ , ] is the Lie bracket on vector fields and $C(u)^C_{AB}$ are known as the Cartan structure functions defined by:
\begin{equation}\text{C}^A_{BC}:=\Gamma^A_{BC}-\Gamma^A_{CB}.\label{CSF}\end{equation}

The corresponding lifted momentum observable \cite{GIMMSY,jdg2} is:
\begin{equation}
\mathfrak{J}_A=\mathfrak{i}_{\mathcal{M}\pi,J^1\pi^\star}^* \circ \tau^* (\Hat{\theta}_A\hook z)=E_A^ip_i^\mu dy_\mu=:p_A^\mu dy_\mu \label{CovNC}
\end{equation}
We choose new coordinates on $J^1\pi^\star$, viz. $(y^\mu,u^i,p_A^\mu)$.
We shall also have occasion to use $\theta^A=E^A_idu^i$. In terms of these new coordinates the vertical multisymplectic potential takes the form:
\begin{equation}
\Theta^V=p_A^\mu{\Hat{\theta}}^A\wedge dy_\mu
\end{equation}
Then the vertical multisymplectic form may be written as:
\begin{equation}
\Omega^V={\Hat{\theta}}^A \wedge dp_A^\mu\wedge dy_\mu +\tfrac{1}{2}p_A^\alpha \text{C}^A_{BC}{\Hat{\theta}}^B \wedge{\Hat{\theta}}^C\wedge dy_\mu
\end{equation}
The Hamiltonian vector field $\mathfrak{X}[\mathfrak{J}_A]$, solving the multisymplectic structural equation, 
\begin{equation}
\mathfrak{X}[\mathfrak{J}_A]\hook\Omega^V=d^V\mathfrak{J}_A
\end{equation}
is given by:
\begin{equation}
\mathfrak{X}[\mathfrak{J}_D]={\Hat{\theta}}_D +p_E^\nu \text{C}^E_{DG}\frac{\partial}{\partial p_G^\nu}
\end{equation} The vertical exterior derivative is d$^V:=d u^i \frac{\partial}{\partial u^i} +d p_i^\mu\frac{\partial}{\partial p_i^\mu}$.
By a short computation we then have:\newline
\begin{boxedminipage}[t]{12.75cm}
{\thm \label{T1}The Algebra generated by the covariant Noether currents $\mathfrak{J}_A$ is non-Abelian and is given by:}
\[
\{\mathfrak{J}_B ,\mathfrak{J}_C\}=\text{C}^A_{BC}\mathfrak{J}_A
\]where $\{\mathfrak{J}_B ,\mathfrak{J}_C\}=\mathfrak{X}[\mathfrak{J}_B]\hook d^V \mathfrak{J}_C$.
\end{boxedminipage}

{\rmk  We argued earlier in this section that the ambient diffeomorphism invariance of first order embedded submanifolds could be encoded into the covariant Noether currents (\ref{CovNC}), since they give a representation of the Lie algebra of the 1-parameter diffeomorphisms generated by an arbitrary choice of the Cartan moving frame, or equivalently to an equivalence class in D(M)$\cong\tfrac{\text{aut} {\mathcal{F}}\text{(M)}}{\text{gau}{\mathcal F}\text{(M)}}$. Theorem \ref{T1} expresses the fact that the covariant Noether currents corresponding to the ambient diffeomorphism symmetry \[Diff(M):M \times N\ \longrightarrow M \times N \quad :\quad : \quad \Phi:(u(y),y)\longrightarrow (\Phi u(y),y)\] acting by bundle automorphisms over the identity on $\Sigma$ satisfy a non-Abelian algebra.
}
{\rmk \label{R2} Note that the Cartan structure functions are defined as the antisymmetrised connections in the real analytic anholonomic basis. We may therefore locally choose a synchronous frame in which the connections and therefore the Cartan structure  functions  vanish. In this way we see that the deformation algebra of Theorem \ref{T1} may be locally Abelianised. We shall see below that the particular form of the structure functions needed to obtain the algebra of Witten (\ref{s2})- (\ref{s4}) also depends on a special choice of local basis, viz. a Darboux basis.}

{\section{The covariant phase space of pseudoholomorphic maps}}

We now turn our attention to a particular example of first order embeddings, viz. the pseudoholomophic embeddings of Riemann surfaces $(\Sigma,\epsilon)$ with complex structure $\epsilon$ into an almost K\"{a}hler manifold with metric $g$, compatible almost complex structure J, and symplectic structure $\omega$, viz. $(M,\omega,g,J)$. Following the discussion in the last section we then have:

{\defn The {\it configuration bundle } $\pi^{\Sigma}$ for the embeddings of Riemann surfaces into a symplectic manifold is the triple \(( M \times \Sigma,\pi^{\Sigma},\Sigma)\).
}

The kinematical phase space is a particular subbundle of the vertical covariant canonical phase space of embeddings. In order to explicitly construct this subbundle we shall require an explicit coordinisation. 

In order to obtain coordinates with the correct dimensionality we now introduce
the J-pseudoholomorphic anholonomic basis vielbeins on M, viz. (E$^a_i$,E$^{a*}_i$) and also vielbeins taking us from the analytic basis  indices $\alpha$ on $\Sigma$ to  the complex analytic holonomic basis indices $(\kappa,\kappa*)$, viz. $(E_\alpha^\kappa,E_\alpha^{\kappa*})$. See {\scshape Appendix I} for more details.
In terms of these veilbeins one then has the identities:
\begin{alignat}{2} \label{multi-momentum identities}
p^\alpha_i&=\text{E}^{a}_ip_{a}^{\kappa}\text{E}_{\kappa}^\alpha &+
\text{E}^{a*}_ip_{a*}^{\kappa}\text{E}_{\kappa}^\alpha &+
\text{E}^{a}_ip_{a}^{\kappa*}\text{E}_{\kappa*}^\alpha +
\text{E}^{a*}_ip_{a*}^{\kappa*}\text{E}_{\kappa*}^\alpha \\ 
p_{a}^{\kappa} &= \text{E}_{a}^ip_i^\alpha \text{E}_\alpha^{\kappa} &\quad  
  p_{a*}^{\kappa} &= \text{E}_{a*}^ip_i^\alpha \text{E}_\alpha^{\kappa} \\
p_{a}^{\kappa*} &= \text{E}_{a}^ip_i^\alpha \text{E}_\alpha^{\kappa*} &\quad  
  p_{a*}^{\kappa*} &= \text{E}_{a*}^ip_i^\alpha \text{E}_\alpha^{\kappa*} 
\end{alignat}
The vielbeins have be so constructed that they are infact eigenvectors of the respective complex structures (see \cite{Hsuing}). That is,
\begin{alignat}{2}
J^i_jE^j_a & =iE^i_a &\quad    J^i_jE^j_{a*} & =-iE^i_{a*} \\
J_i^jE_j^a & =iE_i^a &\quad    J_i^jE_j^{a*} & =-iE_i^{a*} \\
\varepsilon^\alpha_\beta E^\beta_{\kappa}&=iE^\alpha_{\kappa} &\quad   
\varepsilon^\alpha_\beta E^\beta_{\kappa*}&=-iE^\alpha_{\kappa*} \\
\varepsilon_\alpha^\beta E_\beta^{\kappa}&=iE_\alpha^{\kappa} &\quad  
\varepsilon_\alpha^\beta E_\beta^{\kappa*}&=-iE_\alpha^{\kappa*} 
\end{alignat}
We then have the following:\newline
\begin{boxedminipage}[t]{12.75cm}
{\prop\label{Embb}
If $p_i^\alpha$ satisfies the pseudoholomorphicity condition $J^i_jp_i^\alpha = \varepsilon^\alpha_\beta p_j^\beta$ then:
\begin{gather}
p_{a}^{\kappa*}=0 \text{ and } p_{a*}^{\kappa}=0
\end{gather}
}
\end{boxedminipage}

\begin{proof}
If $p_i^\alpha$ satisfies the pseudoholomorphicity condition then:
\begin{equation}
\begin{split}
p_{a*}^{\kappa} &= \text{E}_{a*}^ip_i^\alpha \text{E}_\alpha^{\kappa} \\
&= \text{E}_{a*}^i[-J_i^j \varepsilon^\alpha_\beta p_j^\beta ] \text{E}_\alpha^{\kappa} \\
&= -(i)(-i)\text{E}_{a*}^ip_i^\alpha \text{E}_\alpha^{\kappa} \\
\text{then } p_{a*}^{\kappa}&=-p_{a*}^{\kappa} \\
\Longrightarrow p_{a*}^{\kappa}&= 0 
\end{split}
\end{equation}
That $p_{a}^{\kappa*}=0$ follows by complex conjugation.
\end{proof}

{\rmk One may easily verify that no constraints result from pseudoholomorphicity on $p_{a*}^{\kappa*}$ nor $p_{a}^{\kappa}$.}

Proposition \ref{Embb} enables us to define the subbundle ${\text{J}^1 \pi^*_+{}^\Sigma}$ by the vanishing of half of the components of the multimomenta.

{\defn The subjective submersion $\mathbb{P_+}$ is the fibrewise map:
\begin{equation}
\mathbb{P_+}:{\text{J}^1 \pi^*{}^\Sigma} \longrightarrow {\text{J}^1 \pi^*_+{}^\Sigma} \hspace{5pt} : \hspace{5pt}: \hspace{5pt}
\mathbb{P_+}:(p_i^\alpha)=(p_a^\kappa,  p_{a*}^\kappa ,p_a^{\kappa *} ,p_{a*}^{\kappa *} ) \longrightarrow (p_a^\kappa,0,0,p_{a*}^{\kappa *}) 
\end{equation}
}

The local adapted coordinates on $\pdjb$ may  therefore be written $(x^\alpha,u^i,p_a^\kappa,p_{a*}^{\kappa *})$.  
Let $\mathfrak{i}_{J^1\pi^\star|_+, J^1\pi^\star}:J^1\pi^\star|_+\longrightarrow J^1\pi^\star$ denote the natural embedding  of $J^1\pi^\star|_+$ into $J^1\pi^\star$.

{\defn
The { vertical canonical covariant phase space of pseudoholomorphic maps} is $(\pdjb,{\Omega^V_+})$ where
$\Omega^V_+:= {\mathfrak{i}_{J^1\pi^\star|_+, J^1\pi^\star}}^*[\Omega^{V}].$ }

\section{The induced ambient diffeomorphism deformations  of pseudoholomorphic mappings}
In order to effect the  analysis of the ambient diffeomorphism symmetry of psudoholomorphic embeddings 
we now introduce new vielbeins taking us from the real analytic anholonomic  basis with indices A to the J-pseudoholomorphic anholonomic basis (a,a*), viz $(E_A^a,E_A^{a*})$, see {\scshape Appendix I}.

We may then induce the symmetry generated by the Noether currents and  the multisymplectic potential from $J^1\pi^\star$ to $J^1\pi^\star|_+$ via the pull-back by $\mathfrak{i}_{J^1\pi^\star|_+, J^1\pi^\star}$. The induced objects are then:
\begin{equation}
\begin{split}
\mathfrak{i}_{J^1\pi^\star|_+, J^1\pi^\star}^*(\mathfrak{J}_A)&=
\mathfrak{i}_{J^1\pi^\star|_+, J^1\pi^\star}^*
(E_A^{a}p_{a}^{\kappa} E_{\kappa}^\alpha dx_\alpha + E_A^{a*}p_{a*}^{\kappa*} E_{\kappa*}^\alpha dx_\alpha\\ &+ E_A^{a*}p_{a*}^{\kappa} E_{\kappa}^\alpha dx_\alpha + E_A^{a}p_{a}^{\kappa*} E_{\kappa*}^\alpha dx_\alpha) \\
&=E_A^{a}p_{a}^{\kappa} E_{\kappa}^\alpha dx_\alpha + E_A^{a*}p_{a*}^{\kappa*} E_{\kappa*}^\alpha dx_\alpha
\end{split}
\end{equation}
By a similar calculation one also has:
\begin{equation}
\mathfrak{i}_{J^1\pi^\star|_+, J^1\pi^\star}^*
{\Hat{\theta}}^V = E_A^{a}p_{a}^{\kappa}{\Hat{\theta}}^A \wedge  dx_{\kappa} + E_A^{a*}p_{a*}^{\kappa*}{\Hat{\theta}}^A \wedge  dx_{\kappa*}
\end{equation}
Let us denote the pull-back of $\mathfrak{J}_A$ by $\mathfrak{J}_A|_+$ and the pull-back of $\Theta$ by $\Theta|_+$.
The multisymplectic form corresponding to $\Theta|_+$ is then:
\begin{equation}
\begin{split}
\Omega^V|_+ &= E_A^{a}{\Hat{\theta}}^A \wedge dp_{a}^{\kappa} \wedge  dx_{\kappa}
+ E_A^{a*}{\Hat{\theta}}^A \wedge dp_{a*}^{\kappa*} \wedge  dx_{\kappa*} \\
&- p_{a}^{\kappa}(\omega^{a}_{b}E^{b}_A + \omega^{a}_{b*}E^{b*}_A){\Hat{\theta}}^A \wedge dx_{\kappa} \\
&- p_{a*}^{\kappa*}(\omega^{a*}_{b}E^{b}_A + \omega^{a*}_{b*}E^{b*}_A){\Hat{\theta}}^A \wedge dx_{\kappa*}
\end{split}
\end{equation}
Where $\omega^a_b=\Gamma_{Cb}^a{\Hat{\theta}}^C$ are the corresponding connection 1-forms.
$\mathfrak{J}_A|_+$ is a Hamiltonian form w.r.t.  $\Omega^V|_+$ with Hamiltonian vector field given by:
\begin{equation}
\begin{split}
\mathfrak{X}[\mathfrak{J}_D|_+]= {\Hat{\theta}}_D &+ p_{a}^{\kappa}(\Gamma^{a}_{(Cb)}E^{b}_D  + \Gamma^{a}_{(Cb*)}E^{b*}_D)E^C_{d}\frac{\partial }{\partial p_{d}^{\kappa}} \\
&+ p_{a*}^{\kappa*}(\Gamma^{a*}_{(Cb)}E^{b}_D  + \Gamma^{a*}_{(Cb*)}E^{b*}_D)E^C_{d*}\frac{\partial }{\partial p_{d*}^{\kappa*}} \\
&+ 2p_{a}^{\kappa}(\Gamma^{a}_{[Cb]}E^{b}_D  + \Gamma^{a}_{[Cb*]}E^{b*}_D)E^C_{d}\frac{\partial }{\partial p_{d}^{\kappa}} \\
&+ 2p_{a*}^{\kappa*}(\Gamma^{a*}_{[Cb]}E^{b}_D  + \Gamma^{a*}_{[Cb*]}E^{b*}_D)E^C_{d*}\frac{\partial }{\partial p_{d*}^{\kappa*}} 
\end{split}
\end{equation}
By a rather lengthy, but straightforward calculation one then has the following result:\newline
\begin{boxedminipage}[t]{12.75cm}

 {\thm \label{T2}The Algebra generated by the covariant Noether currents $\mathfrak{J}_A|_+$ is non-Abelian and is given by:}
\[
\{ \mathfrak{J}_B|_+ ,\mathfrak{J}_C|_+\}=\text{C}^A_{BC}\mathfrak{J}_A|_+
\]

\end{boxedminipage}

{\rmk Theorem \ref{T2} shows us that the non-Abelian deformation diffeomorphism symmetry of  embeddings is inherited by the subclass of pseudoholomorphic embeddings. The origin of the remarks of Witten describing his construction of the algebra (\ref{s2})-(\ref{s4}) that the algebra must be amended in order to be consistent with the ``duality constraints'' (\ref{duality1}) and (\ref{duality2}) would thus seem to be at variance with the point of view taken  here.
However, if one works locally then the algebra of deformations can be Abelianised by choosing an synchronous frame (see Remark {\ref{R2}}). It was shown in an earlier version of this preprint that one can still obtain Witten's algebra from the local Abelian algebra if one considers the algebra induced by an extrinsic embedding of $\pdjb$ into $\djb$. The approach originally taken by Witten thus seems to be this extrinsic point of view.  The intrinsic point of view taken in this article has the advantage of: (i) clearly elucidating the origin of the symmetry in Diff(M), (ii) does not make use of an overdetermined coordinate system. }

\section{The graded canonical kinematical phase space of pseudoholomorphic embeddings}

In this section we shall briefly describe the construction of the graded canonical kinematical phase space of pseudoholomorphic embeddings as follows from the general covariant Hamiltonian BRST formalism developed in \cite{jdg2}. In order to contain this section within a reasonable length we have assumed that the reader will have some familiarity this reference.

Consider the natural projection $\iota: M \times \Sigma\longrightarrow M$. We may use this natural projection to pull-back the linear frame bundle L$\mathcal{F}$(M) onto $M \times \Sigma$. One may then construct the graded bundles of the covariant Hamiltonian BRST formalism according to  Figure \ref{F3} (reading from right to left):
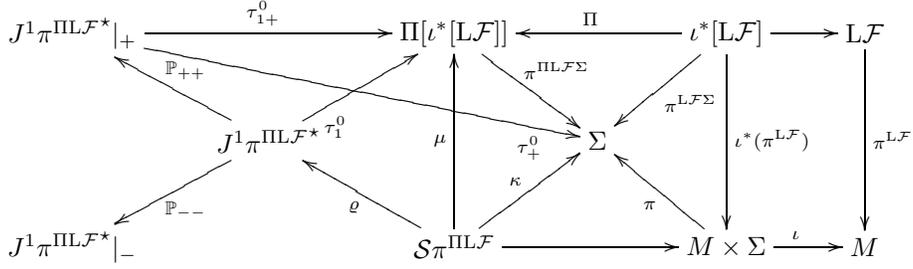
\begin{figure}[h]
\[
\xymatrix{J^1{\pi^{\Pi\text{L}\mathcal{F}}}^\star|_{+} \ar[rr]^{\tau^0_{1+}} \ar[rrrd]_>>>>>>>{\tau^0_+} & &\Pi[\iota^*[\text{L}\mathcal{F}]] \ar[dr]^{\pi^{\Pi\text{L}\mathcal{F}\Sigma}} & & \ar[ll]_\Pi \iota^*[\text{L}\mathcal{F}] \ar[dl]^{\pi^{\text{L}\mathcal{F}\Sigma}} \ar[dd]^{\iota^*({\pi^{\text{L}\mathcal{F}}})} \ar[r]& \text{L}\mathcal{F}\ar[dd]^{\pi^{\text{L}\mathcal{F}}} \\
& \ar[ul]_{\mathbb{P}_{++}}  \ar[dl]^{\mathbb{P}_{--}} J^1{\pi^{\Pi\text{L}\mathcal{F}}}^\star \ar[ur]_<<<<{\tau^0_1} & & \Sigma & & \\
J^1{\pi^{\Pi\text{L}\mathcal{F}}}^\star|_{-} &  & \ar[ul]^\varrho \mathcal{S}\pi^{\Pi\text{L}\mathcal{F}} \ar[ur]^\kappa \ar[uu]^\mu \ar[rr]& &M\times \Sigma \ar[ul]^\pi \ar[r]^\iota& M 
}\]\caption{The construction of the graded bundles.}\label{F3}
\end{figure}
The triple $(J^1{\pi^{\Pi\text{L}\mathcal{F}}}^\star|_{+},\tau^0_{+},\Sigma)$ is the sought after {\it graded vertical covariant canonical  phase space of pseudoholomorphic embeddings}. The construction is outlined as follows (see \cite{jdg2} for the general construction):
\begin{itemize}
\item First construct the pull-back of the frame bundle as a bundle over $M\times\Sigma$ by the natural projection.
\item Then construct the {\it Legendre bundle} $(\iota^*[\text{L}\mathcal{F}],\pi^{\text{L}\mathcal{F}\Sigma},\Sigma)$ where $\pi^{\text{L}\mathcal{F}\Sigma}:=\pi \circ\iota^*(\pi^{\text{L}\mathcal{F}})$.
\item The {\it graded configuration bundle} is then given by the inversion functor (see \cite{voronov}) to be $(\iota^*[\text{L}\mathcal{F}],\pi^{\Pi\text{L}\mathcal{F}\Sigma},\Sigma)$. The adapted local coordinates are $(x^\alpha,u^i,\eta^A)$. The $\eta^A$ carries an $\mathfrak{so}$(m) index and is Grassmann-odd.
\item One then constructs the {\it graded multiphase space } $\mathcal{S}\pi^{\Pi\text{L}\mathcal{F}}$ over the graded configuration bundle according to the usual method. It is endowed with the tautologous canonical form:
\begin{equation*}
{\Theta}^{\text{E}}=p\wedge d^2x + p_i^\alpha du^i \wedge dx_\alpha + \mathcal{P}_A^\alpha d \eta^A \wedge dx_\alpha
\end{equation*}
\item The {\it graded  covariant canonical  phase space bundle of pseudoholomorphic embeddings} is then defined by the  short exact sequence to be $J^1{\pi^{\Pi\text{L}\mathcal{F}}}^\star$. It has local adapted coordinates $(x^\alpha,u^i,\eta^A,p_i^\alpha,\mathcal{P}_A^\alpha)$.
\end{itemize}

Then just as in equation (\ref{multi-momentum identities}) we also have:
\begin{alignat}{2} \label{graded multi-momentum identities}
\mathcal{P}^\alpha_A&=\text{E}^{a}_A\mathcal{P}_{a}^{\kappa}\text{E}_{\kappa}^\alpha &+
\text{E}^{a*}_A\mathcal{P}_{a*}^{\kappa}\text{E}_{\kappa}^\alpha &+
\text{E}^{a}_A\mathcal{P}_{a}^{\kappa*}\text{E}_{\kappa*}^\alpha +
\text{E}^{a*}_A\mathcal{P}_{a*}^{\kappa*}\text{E}_{\kappa*}^\alpha \\ 
\mathcal{P}_{a}^{\kappa} &= \text{E}_{a}^A\mathcal{P}_A^\alpha \text{E}_\alpha^{\kappa} &\quad  
  \mathcal{P}_{a*}^{\kappa} &= \text{E}_{a*}^A\mathcal{P}_i^\alpha \text{E}_\alpha^{\kappa} \\
\mathcal{P}_{a}^{\kappa*} &= \text{E}_{a}^A\mathcal{P}_A^\alpha \text{E}_\alpha^{\kappa*} &\quad  
  \mathcal{P}_{a*}^{\kappa*} &= \text{E}_{a*}^A\mathcal{P}_A^\alpha \text{E}_\alpha^{\kappa*} 
\end{alignat}
in terms of the ``combined vielbeins'', see {\scshape Appendix I}. Moreover,\newline
\begin{boxedminipage}[t]{12.75cm}
{\prop
If $p_i^\alpha$ and $\mathcal{P}^\alpha_A$ satisfy the pseudoholomorphicity conditions $J^i_jp_i^\alpha = \varepsilon^\alpha_\beta p_j^\beta$ and $J^A_B\mathcal{P}_A^\alpha = \varepsilon^\alpha_\beta \mathcal{P}_B^\beta$  then:
\begin{gather}
p_{a}^{\kappa*}=0 \text{ ; } p_{a*}^{\kappa}=0\text{ ; }\mathcal{P}_{a}^{\kappa*}=0 \text{ and } \mathcal{P}_{a*}^{\kappa}=0 
\end{gather}
}
\end{boxedminipage}

{\rmk As before, one may easily verify that no constraints result from pseudoholomorphicity on $p_{a*}^{\kappa*}$, $p_{a}^{\kappa}$, $\mathcal{P}_{a*}^{\kappa*}$ nor on  $\mathcal{P}_{a}^{\kappa}$.}

{\defn The subjective submersion $\mathbb{P_{++}}$ is the fibrewise map:
\begin{equation}
\begin{split}
\mathbb{P}_{++}&:J^1{\pi^{\Pi\text{L}\mathcal{F}}}^\star  \longrightarrow J^1{\pi^{\Pi\text{L}\mathcal{F}}}^\star|_+ \hspace{5pt} : \hspace{5pt}: \hspace{5pt} \\
\mathbb{P}_{++}&:(p_i^\alpha,\mathcal{P}^\alpha_A)=(p_a^\kappa,  p_{a*}^\kappa ,p_a^{\kappa *} ,p_{a*}^{\kappa *},\mathcal{P}_a^\kappa,  \mathcal{P}_{a*}^\kappa ,\mathcal{P}_a^{\kappa *} ,\mathcal{P}_{a*}^{\kappa *} ) \longrightarrow (p_a^\kappa,0,0,p_{a*}^{\kappa *},\mathcal{P}_a^\kappa,0,0,\mathcal{P}_{a*}^{\kappa *}) 
\end{split}
\end{equation}
}
The local adapted coordinates on $J^1{\pi^{\Pi\text{L}\mathcal{F}}}^\star|_+$ may  therefore be written in the following form: $(x^\alpha,u^i,\eta^A, p_a^\kappa,p_{a*}^{\kappa *},\mathcal{P}_a^\kappa,\mathcal{P}_{a*}^{\kappa *})$. 

{\defn The graded extended covariant kinematical phase space of embeddings is the quadruple $(J^1{\pi^{\Pi\text{L}\mathcal{F}}}^\star|_{+},\tau^0_{+},\Sigma,\Omega^V|_+)$ where $\Omega^{\text{EV}}|_+:=(\mathbb{P}_{++} \circ \varrho )^*[-d\Theta^{\text{EV}}]$.
}

In terms of the local adapted coordinates on $J^1{\pi^{\Pi\text{L}\mathcal{F}}}^\star|_+$ one has the following local expression for the induced vertical Cartan form:
\begin{equation}
\begin{split}
\Omega^{\text{EV}}|_+ &= E_A^{a}{\Hat{\theta}}^A \wedge dp_{a}^{\kappa} \wedge  dx_{\kappa}
+ E_A^{a*}{\Hat{\theta}}^A \wedge dp_{a*}^{\kappa*} \wedge  dx_{\kappa*} \\
&-E_A^{a}d\eta^A \wedge d\mathcal{P}_{a}^{\kappa} \wedge  dx_{\kappa}
-E_A^{a*}d\eta^A \wedge d\mathcal{P}_{a*}^{\kappa*} \wedge  dx_{\kappa*} \\
&- p_{a}^{\kappa}(\omega^{a}_{b}E^{b}_A + \omega^{a}_{b*}E^{b*}_A) \wedge {\Hat{\theta}}^A \wedge dx_{\kappa} \\
&- p_{a*}^{\kappa*}(\omega^{a*}_{b}E^{b}_A + \omega^{a*}_{b*}E^{b*}_A) \wedge{\Hat{\theta}}^A \wedge dx_{\kappa*}\\
&+ \mathcal{P}_{a}^{\kappa}(\omega^{a}_{b}E^{b}_A + \omega^{a}_{b*}E^{b*}_A)\wedge d\eta^A \wedge dx_{\kappa} \\
&+ \mathcal{P}_{a*}^{\kappa*}(\omega^{a*}_{b}E^{b}_A + \omega^{a*}_{b*}E^{b*}_A)\wedge d\eta^A \wedge dx_{\kappa*}
\end{split}
\end{equation}

\section{The BRST algebra of the ambient diffeomorphism deformations of pseudoholomorphic embeddings}

Recall that in Theorem \ref{T1} we proved that the algebra of the covariant Noether currents corresponding to the ambient diffeomorphism deformations of embeddings satisfied a non-Abelian algebra whose structure functions are the structure functions of Cartan. That is:
\[
\{ \mathfrak{J}_B ,\mathfrak{J}_C\}=\text{C}^A_{BC}\mathfrak{J}_A
\]
According to the general formalism developed in \cite{jdg2}, the covariant BRST Noether current is the Grassmann-odd 1-form:
\begin{equation}
\Upsilon=\eta^A\mathfrak{J}_A-\tfrac{1}{2}\text{C}^A_{BC}\eta^B\eta^C\mathcal{P}_A^\alpha dx_\alpha
\end{equation}
We shall not pause to consider the BRST algebra generated by $\Upsilon$ but shall pass on to the case of pseudoholomorphic embeddings. The algebra is given, as in Theorem \ref{T2}, by:
\[
\{ \mathfrak{J}_B|_+ ,\mathfrak{J}_C|_+\}=\text{C}^A_{BC}\mathfrak{J}_A|_+
\]
The the covariant BRST Noether current is the Grassmann-odd 1-form corresponding to this  algebra is the pull back $\Upsilon|_+=(\mathbb{P}_{++}\circ\varrho)^*[\Upsilon]$ which is given by:
\begin{equation}
\Upsilon|_+=\eta^A\mathfrak{J}_A|_+ -\tfrac{1}{2}\text{C}^A_{BC}\eta^B\eta^C\mathcal{P}_A^\alpha|_+ dx_\alpha
\end{equation}
where $\mathcal{P}_A^\alpha|_+=\text{E}^{a}_A\mathcal{P}_{a}^{\kappa}\text{E}_{\kappa}^\alpha  +
\text{E}^{a*}_A\mathcal{P}_{a*}^{\kappa*}\text{E}_{\kappa*}^\alpha$.
In order to calculate the BRST algebra we must first calculate the Hamiltonian vector fields corresponding to the observables, which are to be built from the ``field content'' of the theory. The ``fields'' in this case are the local fibre coordinates on $(J^1{\pi^{\Pi\text{L}\mathcal{F}}}^\star|_{+},\tau^0_{+},\Sigma,\Omega^V|_+)$.That is, the field content is: \[(u^i,\eta^A, p_a^\kappa,p_{a*}^{\kappa *},\mathcal{P}_a^\kappa,\mathcal{P}_{a*}^{\kappa *}).\]
We construct from this field multiplet the following Hamiltonian forms:
\begin{alignat}{1}
u^i; &\quad \mathfrak{J}_A|_+; \\
\eta^A; &\quad \mathcal{P}_A|_+.
\end{alignat}
The corresponding Hamiltonian form-valued (multi-)vector fields are:
\begin{equation}
\mathfrak{X}[u^i]=\text{E}^i_{a} \frac{\partial}{\partial z^{\kappa *}} \wedge \frac{\partial}{\partial p_{a}^{\kappa}} + \text{E}^i_{a*} \frac{\partial}{\partial z^{\kappa }} \wedge \frac{\partial}{\partial p_{a*}^{\kappa *}}       
\end{equation}
\begin{equation}
\begin{split}
\mathfrak{X}[\mathfrak{J}_D|_+]= {\Hat{\theta}}_D &+ p_{a}^{\kappa}(\Gamma^{a}_{(Cb)}E^{b}_D  + \Gamma^{a}_{(Cb*)}E^{b*}_D)E^C_{d}\frac{\partial }{\partial p_{d}^{\kappa}} \\
&+ p_{a*}^{\kappa*}(\Gamma^{a*}_{(Cb)}E^{b}_D  + \Gamma^{a*}_{(Cb*)}E^{b*}_D)E^C_{d*}\frac{\partial }{\partial p_{d*}^{\kappa*}} \\
&+ 2p_{a}^{\kappa}(\Gamma^{a}_{[Cb]}E^{b}_D  + \Gamma^{a}_{[Cb*]}E^{b*}_D)E^C_{d}\frac{\partial }{\partial p_{d}^{\kappa}} \\
&+ 2p_{a*}^{\kappa*}(\Gamma^{a*}_{[Cb]}E^{b}_D  + \Gamma^{a*}_{[Cb*]}E^{b*}_D)E^C_{d*}\frac{\partial }{\partial p_{d*}^{\kappa*}} \\
&- \mathcal{P}_{a}^{\kappa}(\Gamma^{a}_{Db}E^{b}_A + \Gamma^{a}_{Db*}E^{b*}_A)d\eta^A \text{E}^E_{e}\frac{\partial}{\partial p_{e}^{\kappa}}\wedge {\Hat{\theta}}_E \\
&- \mathcal{P}_{a*}^{\kappa*}(\Gamma^{a*}_{Db}E^{b}_A + \Gamma^{a*}_{Db*}E^{b*}_A) d\eta^A \text{E}^E_{e*}\frac{\partial}{\partial p_{e*}^{\kappa*}}\wedge {\Hat{\theta}}_E 
\end{split}
\end{equation}
\begin{equation}      
\mathfrak{X}[\eta^A]= - \biggl(\text{E}^A_{a} \frac{\partial}{\partial z^{\kappa *}} \wedge \frac{\partial}{\partial \mathcal{P}_{a}^{\kappa}} + \text{E}^i_{a*} \frac{\partial}{\partial z^{\kappa }} \wedge \frac{\partial}{\partial \mathcal{P}_{a*}^{\kappa *}} \biggr)
\end{equation}
\begin{equation}
\begin{split}
\mathfrak{X}[\mathcal{P}_A|_+]=-\frac{\partial}{\partial \eta^A}
\end{split}
\end{equation}
The BRST algebra is then obtained by taking the interior product of the above 
form-valued (multi-)vector fields with the exterior derivative of $\Upsilon|_+$. The latter object may be expressed as:
\begin{equation} 
\begin{split}
d\Upsilon|_+&= d\eta^A \wedge \mathfrak{J}_A|_+ +\eta^A d \mathfrak{J}_A|_+
-\text{C}^A_{BC}(d\eta^B)\eta^C\mathcal{P}_A^\alpha|_+ dx_\alpha \\
&-\tfrac{1}{2}\text{C}^A_{BC}\eta^B\eta^C d\mathcal{P}_A^\alpha|_+ dx_\alpha
-\tfrac{1}{2}[\Hat{\theta}_D\text{C}^A_{BC}\Hat{\theta}^D]\eta^B\eta^C\mathcal{P}_A^\alpha|_+ dx_\alpha
\end{split}
\end{equation}
By a straightforward calculation we then have:\newline
\begin{boxedminipage}[t]{12.75cm}

{\thm[Ambient diffeo BRST Algebra: the pseudoholomorphic case] \label{T3} The BRST algebra corresponding to the diffeomorphism deformation algebra of pseudoholomorphic embeddings is given by:}
\begin{align}
\delta_{\Upsilon|_+}[u^i]&:= \mathfrak{X}[u^i]\hook d  \Upsilon=\text{E}^i_A\eta^A=:\chi^i                         \\
\delta_{\Upsilon|_+}[\chi^i]&:=    \mathfrak{X}[\chi^i]\hook d  \Upsilon|_+  =0                           \\
\delta_{\Upsilon|_+}[\mathcal{P}_A|_+]&:=   \mathfrak{X}[\mathcal{P}_A|_+]\hook d  \Upsilon|_+ = - [\mathfrak{J}_A|_+ +
        \text{C}^B_{AC}\eta^C\mathcal{P}_B^\alpha|_+ dx_\alpha         ] \\
\delta_{\Upsilon|_+}[\mathfrak{J}_A|_+]&:=    \mathfrak{X}[\mathfrak{J}_A|_+]\hook d  \Upsilon|_+= \eta^C \text{C}^D_{AC}\mathfrak{J}_D|_+
  - \tfrac{1}{2}[\text{D}_A\text{C}^D_{BC}]\eta^B\eta^C\mathcal{P}_D^\alpha|_+ dx_\alpha \label{LT}  \end{align} 

\end{boxedminipage}

The covariant derivative appears in the last term of the transformation (\ref{LT}) by virtue of identity (\ref{id12}) in Appendix II.

\section{Relation to Witten's BRST algebra}

We can already see the beginnings of an emergence of Witten's algebra in Theorem \ref{T3}. What is missing is the fact that the structure functions that are implicit in the construction of Witten involve the covariant derivatives of the almost complex structures. This form of the structure functions, as we shall show below, are in fact the Cartan structure functions expressed in a Darboux basis on M. They are therefore an artifact of the local geometry of the symplectic manifold  M. Once we establish this fact the algebra of Witten then follows.\newline
\begin{boxedminipage}[t]{12.75cm}

{\prop \label{P1} The Levi-Civita connection on a symplectic manifold, expressed  in terms of a real anholonomic Darboux basis is given by the following expression:}
\begin{equation}
\Gamma^A_{BC}=\tfrac{1}{2}\text{J}^A_D\text{D}_B\text{J}^D_C\label{E3}
\end{equation}

\end{boxedminipage}

\begin{proof}
We perform all computations in a real anholonomic Darboux basis.
Consider the compatibility equation between the almost complex structure J, symplectic form $\omega$ and metric $g$, viz. \(\omega_{AB}=J^C_Ag_{CB} \). Taking the covariant derivative of both sides and by  use of the compatibility equation we find that:
\begin{equation}
\begin{split}
\text{D}_E\omega_{AB}&=(\text{D}_EJ^C_A)g_{CB} \\
&=-(\text{D}_EJ^C_A)J^F_CJ_F^Gg_{GB}\\
&=-J^F_C(\text{D}_EJ^C_A)\omega_{FB}\label{E1}
\end{split}
\end{equation}
Computing the covariant derivative in local Darboux basis on M:
\begin{equation}
\text{D}_E\omega_{AB}= -2\Gamma^F_{AE}\omega_{FB}\label{E2}
\end{equation}
On comparison of equations (\ref{E1}) and (\ref{E2}) one obtains (\ref{E3}).
\end{proof}
We then obtain the structure functions of Witten by antisymmetrising:\newline
\begin{boxedminipage}[t]{12.75cm}

{\cor The Cartan structure functions in a real anholonomic Darboux basis  on M are identical to Witten's structure functions, viz. }
\begin{equation}
\text{C}^A_{BC}:=\Gamma^A_{[BC]}=\tfrac{1}{2}\text{J}^A_D\text{D}_{[B}\text{J}^D_{C]}\label{E4}
\end{equation}
\end{boxedminipage}

Together with Theorem \ref{T3}, this Corollary then yields Witten's BRST algebra:\newline
\begin{boxedminipage}[t]{12.75cm}
{\thm[Witten's Algebra] \label{T5} 
The BRST algebra corresponding to the diffeomorphism deformation algebra of pseudoholomorphic embeddings in a real anholonomic Darboux basis   is given by:}
\begin{align}
\delta_{\Upsilon|_+}[u^i]&:= \mathfrak{X}[u^i]\hook d  \Upsilon=\text{E}^i_A\eta^A=:\chi^i                         \\
\delta_{\Upsilon|_+}[\chi^i]&:=    \mathfrak{X}[\chi^i]\hook d  \Upsilon|_+  =0                           \\
\delta_{\Upsilon|_+}[\mathcal{P}_A|_+]&:=   \mathfrak{X}[\mathcal{P}_A|_+]\hook d  \Upsilon|_+ = - [\mathfrak{J}_A|_+ +
        \tfrac{1}{2}\text{J}^B_D\text{D}_{[A}\text{J}^D_{C]}\eta^C\mathcal{P}_B^\alpha|_+ dx_\alpha         ] \\
\delta_{\Upsilon|_+}[\mathfrak{J}_A|_+]&:=    \mathfrak{X}[\mathfrak{J}_A|_+]\hook d  \Upsilon|_+= \eta^C \tfrac{1}{2}\text{J}^D_F\text{D}_{[A}\text{J}^F_{C]}\mathfrak{J}_D|_+ \\
  &- \tfrac{1}{4} [(\text{D}_{A} J^D_F ) (\text{D}_{B}\text{J}_{C}^F) 
+ J^D_F\text{R}_{AB}{}^{\text{  } F}{}_{\text{}K}\text{J}_{C}^K
+ J^D_F\text{R}_{CB}J^F_{A}]\eta^B\eta^C\mathcal{P}_D^\alpha|_+ dx_\alpha  \tag*{}        \end{align} 

\end{boxedminipage}

\begin{proof}
The first three equations are obtained from those of Theorem \ref{T3} by direct substitution from equation (\ref{E4}). The final equation follows by substitution and by identity (\ref{id13}) from Appendix II.\end{proof} 
{\rmk $\chi^i$ is to be identified with the anticommuting parameter introduced by Witten. We now see that the commuting field H$^{i\alpha}$ is the multimomenta p$_i^\alpha$ with one index raised, and the anticommuting field $\varrho^\alpha_i$ is the ghost multimomenta $\mathcal{P}^\alpha_i$.   }

{ \rmk The extra {\it ``...non-covariant looking terms...''} which Witten \cite{TSM} added to his transformations do not occur here because we know the \(p_i^\alpha|_{+}\) and the \(\mathcal{P}_A^\alpha|_{+}\) are geometrically local fibre coordinates on an affine bundle. They are therefore expected to transform affinely under reparametrisations of the \(u^i\). We have therefore achieved the aim of this section and given the first  derivation of this BRST algebra in any Hamiltonian formulation of BRST.}

\newpage

\section*{Appendix I: Vielbein Identities}

We shall have use of three  sets of vielbeins:
\begin{itemize}
\item Real analytic anholonomic basis vielbeins on M.
\item J-pseudoholomorphic anholonomic basis vielbeins on M. 
\item Complex analytic holonomic basis vielbeins on $\Sigma$.
\end{itemize}
We collect together here the identities satisfied by these  vielbeins.

We denote the indices of the anholonomic orthonormal vielbein basis (for more details see \cite{Nakahara}) on M by upper case Latin letters:  $A,B,\cdots=1,\cdots,m$ and the coordinate basis by lower case middle Latin letters: $i,j=1,\cdots,m$. They satisfy the following identities:

{\hspace{5pt}}\newline
\begin{boxedminipage}[t]{12.75cm}

{ \scshape Real analytic anholonomic basis vielbeins on M:}

\begin{alignat}{2}
\text{E}_A^i\text{E}_i^B  &= \delta^B_A  &\qquad\text{E}_A^i\text{E}_j^A &= \delta^i_j \tag{RA1}\\
\Hat{\theta}^A &= \text{E}_j^A du^j &\qquad  \Hat{\theta}_A &=\text{E}_A^i\frac{\partial}{\partial u^i}\tag{RA2} \\
g(\Hat{\theta}_A,\Hat{\theta}_B):&=\delta_{AB}  &\qquad g_{ij} &=\text{E}_A^i\text{E}_B^j\delta_{AB} \tag{RA3}
\end{alignat}
\end{boxedminipage}

{\hspace{5pt}}\newline
The following characteristic  of almost complex manifolds may be found in  in \cite{Hsuing}:
{\thm  A necessary and sufficient condition for a differentiable 2k-dimensional (2k=m) manifold $M^{\text{2k}}$ to admit an almost complex structure is that on $M^{\text{2k}}$ there exist two distributions $\Pi_{k}$ and $\Bar{\Pi}_{k}$ of complex dimension k, which are conjugate to each other, have no common direction and span a linear space of dimension 2n.}

So by this theorem let  \(( \text{E}^i_a )_{a=1,\cdots,k}\) span $\Pi_{k}$ and \( (\text{E} ^i_{a*} )_{a*=1,\cdots,k}\) span  $\Bar{\Pi}_{k}$. Since the 2n vectors \( ( \text{E}^i_a, \text{E}^i_{a*} )_{a,a*=1,\cdots,k}\) are also linearly independent, the matrix \( ( \text{E}^i_a, \text{E}^j_{a*} )\) has an inverse which is denoted by \( ( \text{E}_i^a, \text{E}_j^{a*} )\).
Then the following hold:

{\hspace{5pt}}\newline
\begin{boxedminipage}[t]{12.75cm}

{ \scshape J-pseudoholomorphic anholonomic basis vielbeins on M:}

\begin{alignat}{2}
\text{E}_{a}^i\text{E}_i^{b}  &= \delta^{a}_{b}  &\qquad  \text{E}_{a*}^i\text{E}_i^{b*} &= \delta^{a*}_{b*}\tag{JA1}\\
\text{E}_{a}^i\text{E}_i^{b*}  &= 0  &\qquad  \text{E}_{a*}^i\text{E}_i^{b} &= 0\tag{JA2}\\
\text{E}_{a}^i\text{E}_j^{a} + \text{E}_{a*}^i\text{E}_j^{a*} &=\delta^i_j 
\tag{JA3}\\
\Hat{\theta}^a &= \text{E}_j^a du^j &\qquad  \Hat{\theta}^{a*} &= \text{E}_j^{a*} du^j \tag{JA4}\\
\Hat{\theta}_{a} &=\text{E}_{a}^i\frac{\partial}{\partial u^i}  &\qquad  
\Hat{\theta}_{a*} &=\text{E}_{a*}^i\frac{\partial}{\partial u^i}\tag{JA5}
\end{alignat}

\end{boxedminipage}

For more details see \cite{Hsuing,Goldberg}.
\newpage

Since the almost complex structure on $\Sigma$ is actually  integrable we have the following holonomic vielbeins corresponding to the complex coordinate basis.

{\hspace{5pt}}\newline
\begin{boxedminipage}[t]{12.75cm}

{ \scshape Complex analytic holonomic basis vielbeins on $\Sigma$:}

\begin{alignat}{2}
\text{E}_{\kappa}^\alpha\text{E}_\alpha^{\lambda}  &= \delta_{\kappa}^{\lambda}  &\qquad  \text{E}_{\kappa*}^\alpha\text{E}_\alpha^{\lambda*} &= \delta^{\lambda*}_{\kappa*}\tag{CH1}\\
\text{E}_{\kappa}^\alpha\text{E}_\alpha^{\lambda*}  &= 0  &\qquad  \text{E}_{\kappa*}^\alpha\text{E}_\alpha^{\lambda} &= 0\tag{CH2}\\
\text{E}_{\kappa}^\alpha\text{E}_\beta^{\kappa} + \text{E}_{\kappa*}^\alpha\text{E}_\beta^{\kappa*} &=\delta^\alpha_\beta \tag{CH3}\\
\Hat{\theta}^\kappa &= \text{E}_\alpha^\kappa dx^\alpha=dz^\kappa &\qquad  \Hat{\theta}^{\kappa*} &= \text{E}_\alpha^{\kappa*} du^\alpha=dz^{\kappa*}\tag{CH4} \\
\Hat{\theta}_{\kappa} &=\text{E}_{\kappa}^\alpha\frac{\partial}{\partial x^\alpha}= \frac{\partial}{\partial z^\kappa} &\qquad  
\Hat{\theta}_{\kappa*} &=\text{E}_{\kappa*}^\alpha\frac{\partial}{\partial x^\alpha}=\frac{\partial}{\partial z^{\kappa*} }\tag{CH5}
\end{alignat}
\vspace{5pt}

\end{boxedminipage}

\vspace{10pt}

We shall also need to make use of the following ``combination vielbeins'' on M,
defined by $\text{E}_{a}^A:=\text{E}_{a}^i\text{E}_{i}^A$ and satisfying the following identities:

{\hspace{5pt}}\newline
\begin{boxedminipage}[t]{12.75cm}

{ \scshape ``combination'' anholonomic basis vielbeins on M:}

\begin{alignat}{2}
\text{E}_{a}^A\text{E}_A^{b}  &= \delta^{a}_{b}  &\qquad  \text{E}_{a*}^A \text{E}_A^{b*} &= \delta^{a*}_{b*}\tag{CBA1}\\
\text{E}_{a}^A\text{E}_A^{b*}  &= 0  &\qquad  \text{E}_{a*}^A \text{E}_A^{b} &=0\tag{CBA2} \\
\text{E}_{a}^A\text{E}_B^{a} + \text{E}_{a*}^A\text{E}_B^{a*} &=\delta^A_B \tag{CBA3}
\end{alignat}

\end{boxedminipage}

\newpage

\section*{Appendix II:On almost K\"ahler  manifolds}

We shall give a brief review and some useful geometric  identities satisfied by  of almost K\"ahler structures. Almost  K\"ahler manifolds are one class of a richly structured interrelated hierarchy  of manifolds. Only the K\"ahler class has drawn much attention from the theoretical physics community, via the pre-eminence of Calabi-Yau manifolds in string theory\footnote{Although the almost K\"ahler class, in the form of symplectic manifolds, is the axiomatic foundation of classical mechanics and plays a indispensable role in geometric quantisation.}. This  hierarchy\footnote{For more details see \cite{Hsuing}. } may be depicted graphically as in Figure \ref{K}:
\begin{figure}[h]
\[
\xymatrix{ & \mathcal{AK} \ar@0[dr]^{{\subset}}\\
 \mathcal{K} \ar@0[ur]^\subset \ar@0[dr]_\subset& &\mathcal{QK} \hspace{10pt} \subset & \mathcal{ASK}\hspace{10pt} \subset & \mathcal{AH} \\
 & \mathcal{NK} \ar@0[ur]_\subset
}
\]\caption{The hierarchy of almost K\"ahler  manifolds.}\label{K}
\end{figure}
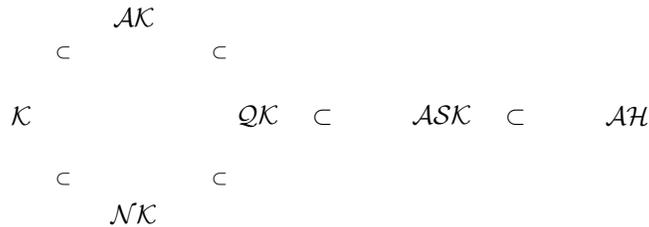
\index{almost Hermitian manifold}\index{almost semi- K\"ahler manifolds manifold}\index{quasi-K\"ahler manifold}\index{nearly K\"ahler manifold}\index{almost K\"ahler  manifold}\index{K\"ahler  manifold} where \(\mathcal{K}\) denotes K\"ahler, \(\mathcal{AK}\) denotes almost K\"ahler, \(\mathcal{NK}\) denotes nearly K\"ahler, \(\mathcal{QK}\) denotes quasi-K\"ahler,  \(\mathcal{ASK}\) denotes almost semi-K\"ahler and  \(\mathcal{AH}\) denotes almost Hermitian.
All almost complex manifolds admit a Hermitian metric, and they are thus all {\it almost Hermitian}. The next most general class are the {\it almost semi- K\"ahler manifolds}. They are defined to be those almost Hermitian manifolds whose almost complex structure obeys the identity: 
\begin{equation}
\text{J}_i:=-\text{D}_k\text{J}_i^k=0\label{id1} \tag{A.1}
\end{equation}
One then obtains the {\it quasi-K\"ahler} manifolds by imposing the identity: 
\begin{equation}
\text{D}_k\text{J}_i^j +\text{J}_r^k\text{J}_s^i\text{D}_r\text{J}_s^j=0\label{id2}\tag{A.2}
\end{equation}

At this point the classification bifurcates into two classes. The { \it almost K\"ahler} class was described in the introduction. The {\it nearly K\"ahler} class is obtained from the quasi-K\"ahler class by imposing the identity \(\text{D}_{[k}\text{J}_{i]}^l=0\). Finally both the almost K\"ahler and nearly K\"ahler classes coalesce to the ``integrable''  {\it K\"ahler} case upon certain extra conditions upon their respective almost complex structures.

Having described the wider context  in which almost K\"ahler manifolds fit we now detail some identities which are satisfied by the corresponding almost complex structure. Knowledge of the wider context will prove to be useful in proving the various identities which follow. 

As a consequence of the fact that the almost complex structure is an anti-idempotent endomorphism of the tangent bundle, viz \( \text{J}^m_n  \text{J}^n_l=-\delta^m_l \), one finds that:
\begin{equation}
\text{J}^m_n \text{D}_k\text{J}^n_l =- \text{J}^n_l \text{D}_k \text{J}^m_n 
\text{   or equivalently,   } \text{J}^m_n \partial_k\text{J}^n_l =- \text{J}^n_l \partial_k \text{J}^m_n \label{id3}\tag{A.3}
\end{equation}
Now consider the tensor \( \text{C}_{mij}:=\tfrac{1}{2}\text{J}_{mn} \text{D}_{[i} \text{J}_{j]}^n \) which plays the role of structure functions for the non-Abelian deformation algebra. 
\begin{equation}
\begin{split}
\text{C}_{mij}&:=\tfrac{1}{2}\text{J}_{mn} \text{D}_{[i} \text{J}_{j]}^n  \\
&=- \tfrac{1}{2}(\text{D}_{[i|}\text{J}_{mn}) \text{J}_{|j]}^n \\
&=-  \tfrac{1}{2} (\text{D}_{[i|} \text{J}_{m}^t) \text{J}_{|j]t} \\ 
&=- \text{C}_{jim}
\end{split}\tag{A.4}
\end{equation}
The tensor \( \text{C}_{mij} \) is therefore antisymmetric in all three indices.
By virtue of  the quasi-K\"ahler identity (\ref{id2}) we have:
\begin{equation}
\begin{split}
\text{C}_{mkl}&=-\tfrac{1}{2}\text{J}^n_m\text{J}^r_{[k|} \text{J}^s_n \text{D}_r\text{J}_{s|l]} \\
&=\tfrac{1}{2}\text{J}^r_{[k|}\text{D}_r\text{J}_{m|l]} \\
&=-\tfrac{1}{2}\text{J}^n_{m}\text{D}_n\text{J}_{[kl]} \\
\Longleftrightarrow  0&=\text{J}_{m}^n \text{D}_{([k|} \text{J}_{l]|n)}
\end{split}\label{id5}\tag{A.5}
\end{equation}
We now demonstrate that the covariant derivative of the structure functions satisfies an antisymmetry. More precisely:
\begin{equation}
\text{D}_q\text{C}_{mkl}=:\tfrac{1}{2}\text{D}_q\{ 
J^n_m\text{D}_{[k|}\text{J}_{n|l]}
\}=\tfrac{1}{2}\text{D}_{[q|}\{   J^n_m\text{D}_{[|k]|}\text{J}_{n|l]}
 \} \label{id6}\tag{A.6}
\end{equation}
By virtue of the Leibniz rule the covariant derivative of the structure functions splits into two parts:
\begin{equation}
\text{D}_q\text{C}_{mkl}= \tfrac{1}{2}(\text{D}_q J^n_m ) (\text{D}_{[k|}\text{J}_{n|l]}) 
+ \tfrac{1}{2}J^n_m(\text{D}_q\text{D}_{[k|}\text{J}_{n|l]} )\label{id7}\tag{A.7}
\end{equation}
We shall establish the antisymmetry of each of these terms and thus of the whole. Consider the first term:
\begin{equation}
\begin{split}
(\text{D}_q J^n_m ) (\text{D}_{[k|}\text{J}_{n|l]})&=
-(\text{D}_q J^n_{[l} ) (\text{D}_{k]}\text{J}_{nm})\\
&=(\text{D}_q J_{[l|n} ) (\text{D}_{|k]}\text{J}_{m}^n) \\
&=-(\text{D}_q J_{n[l} ) (\text{D}_{k]}\text{J}_{m}^n) \\
&= - (\text{D}_{[k|}\text{J}_{m}^n)(\text{D}_q J_{n|l]} )\label{id8}
\end{split}\tag{A.8}
\end{equation}
where we have used the complete antisymmetry of the structure functions and have carefully raised and lowered using the covariantly constant metric on M. Now we consider the second term, the antisymmetry of which will follow by demonstrating that the following is true:
\begin{equation}
J^n_m\text{D}_{(q}\text{D}_{[|k)|}\text{J}_{n|l]}=0\label{id9}\tag{A.9}
\end{equation}
To prove this consider the following:
\begin{equation}
\begin{split}
J^n_m\text{D}_{(q}\text{D}_{[k)|}\text{J}_{n|l]}
&=- J^n_m\text{D}_{(q}\text{D}_{[l|}\text{J}_{n||k])} \\
&=- J^n_m\text{D}_{([k|}\text{D}_{l]}\text{J}_{n|q)} \\
&=  J^n_m\text{D}_{([l|}\text{D}_{k]}\text{J}_{n|q)} \\
&=-  J^n_m\text{D}_{[l|}\text{D}_{n}\text{J}_{|(k]q)} \\  
&= 0 \label{id10}
\end{split}\tag{A.10}
\end{equation}
where the second from last equality follows from (\ref{id5}). We therefore conclude that \( \text{D}_q\text{C}_{mkl}\) is antisymmetric in q and k as required, i.e. \( \text{D}_q\text{C}_{mkl}\) is completely antisymmetric. This also holds in the real anholonomic basis so that \( \text{D}_A\text{C}_{BCD}\) is completely antisymmetric. It follows from the complete antisymmetry of \( \text{D}_A\text{C}_{BCD}\) that:
\begin{equation}
\text{D}_A\text{C}_{BCD}=\Hat{\theta}_A\text{C}_{BCD}\tag{A.11}\label{id11}
\end{equation}
and since the metric in the anholonomic basis is given by \(g_{AB}=\delta_{AB}\)one finds that:
\(
\Hat{\theta}_A\text{C}_{BCD}=(\Hat{\theta}_A\text{C}_{CD}^E)\delta_{EB}
\).
From which it follows:
\begin{equation}
\Hat{\theta}_A\text{C}_{CD}^E=\text{D}_A\text{C}_{CD}^E\tag{A.12}\label{id12}
\end{equation}
By virtue of the Ricci identity, see \cite{Hsuing}, one has  \(\text{J}^m_n[\text{D}_q,\text{D}_k]J^n_l=\text{J}^m_n \text{R}_{qk}{}^{\text{  } n}{}_{\text{}r}\text{J}^r_l+ \text{J}^m_n\text{R}_{lk}J^n_q\). We then find (in the anholonomic basis):
\begin{equation}
\text{D}_A\text{C}_{BC}^D=(\text{D}_{A} J^D_F ) (\text{D}_{B}\text{J}_{C}^F) 
+ J^D_F\text{R}_{AB}{}^{\text{  } F}{}_{\text{}K}\text{J}_{C}^K
+ J^D_F\text{R}_{CB}J^F_{A}\tag{A.13}\label{id13}
\end{equation}

\end{document}